# The covert set-cover problem with application to Network Discovery [*]

Sandeep Sen
Dept of Comp Sc & Engg
IIT Delhi
New Delhi-110016, India
ssen@cse.iitd.ernet.in

V.N. Muralidhara
IIIT-Bangalore,
Electronics City,
Bangalore, 560100, India
murali@iiitb.ac.in

November 15, 2018

## Abstract

We address a version of the set-cover problem where we do not know the sets initially (and hence referred to as covert) but we can query an element to find out which sets contain this element as well as query a set to know the elements. We want to find a small set-cover using a minimal number of such queries. We present a Monte Carlo randomized algorithm that approximates an optimal set-cover of size $OPT$ within $O(\log N)$ factor with high probability using $O(OPT \cdot \log^2 N)$ queries where $N$ is the input size.

We apply this technique to the network discovery problem that involves certifying all the edges and non-edges of an unknown $n$-vertices graph based on layered-graph queries from a minimal number of vertices. By reducing it to the covert set-cover problem we present an $O(\log^2 n)$-competitive Monte Carlo randomized algorithm for the covert version of network discovery problem. The previously best known algorithm [4] has a competitive ratio of $\Omega(\sqrt{n \log n})$ and therefore our result achieves an exponential improvement.

## 1 Introduction

Given a ground set $S$ with $n'$ elements and a family of sets $S_1, S_2 \ldots S_{m'}$ where[1] $S_i \subset S$, a *cover* $C$ is a collection of sets from this family whose union is $S$. It is known that finding a cover consisting of the minimum number of sets is a computationally intractable problem [9]. There are many strategies [6, 11, 14] to *approximate* the smallest cover within a factor of $O(\log n')$ which is known to be the best possible unless $P = NP$ [8].

In this paper, we consider the following version of the set cover problem. Although we know $m', n'$, we do not know the elements nor the cardinality of any of the sets $S_i$. We are allowed to query an element $e \in S$ that returns all sets $S_i$ that contain $e$ which we refer to as a hitting-set query; we can also query a set to know its elements. We would like to compute a small set cover of $S$ using a minimal number of such queries. More specifically, if $OPT$ is the minimum size of a set cover for any instance of the problem, we would like to find a set cover of size $O(OPT \cdot \text{polylog } n')$ using only $O(OPT \cdot \text{polylog } n')$ queries. Note that by using $\min\{m', n'\}$ queries, we can reduce it to the standard version but the number of queries may not satisfy $O(OPT \cdot \text{polylog } n')$. By restricting the number of queries to be close to $OPT$, an algorithm cannot afford to learn the contents of all the sets, yet it is required to find a cover close to the optimal.

---

[*] A Preliminary version of the results have appeared earlier in the 4th Workshop on Algorithms and Computation 2010

[1] We have chosen $n', m'$ as notations to keep them distinct from graphs with $n$ vertices and $m$ edges.

This formalization is also distinct from the *online* problems addressed in [1, 2] where the sets are known but the adversary chooses a set of the ground set for which a minimal cover must be computed. An adversary chooses the elements one after the other and the online algorithm must maintain a cover of the elements revealed upto a given stage. There is no apparent relationship between the two versions. In one case, the initial sets are not known but the algorithm can choose the elements for hitting set queries whereas in the online case, the sets are known but the adversary chooses the elements. Moreover, the number of queries is also a measure of performance in the version considered here.

Our research is motivated by the problem of discovering the topologies of large networks such as the Internet. For large networks such as the Internet which changes frequently, it is very difficult and costly to obtain the topology accurately. Nevertheless, such information about the network is very useful - for example, the robustness properties of the network or studying the routing aspects.

In order to create the topology of the network, one of the techniques used is to obtain local views of the network from various locations and combine them to determine the topology of the network. One can view this technique as an approach for discovering the topology of the network by some queries. Here, a query corresponds to the local view of the network from one specific location. In the real world scenario, the cost of answering a query is usually very high, so the objective of the network discovery problem is to find the map of the network using a minimal number of queries.

Note that in the network discovery problem, we have to confirm the existence and non-existence of an edge between any pair of vertices. So, any query at a vertex should implicitly or explicitly confirm the absence or presence of edges between some pair of vertices. The *Layered Graph Query Model* and *Distance Query Model* are the most widely studied query models.

*Layered Graph Query Model*: A query at a vertex $v$ yields the set of all edges on shortest paths from the vertex $v$ to any other vertex reachable from $v$ in the graph. More specifically, we obtain information about an edge $(x, y)$, iff $d(v, x)$ and $d(v, y)$ are consecutive where $d(v, x)$ is the level of $x$ (from $v$, see Figure 1).

*Distance Query Model*: A query at a vertex $v$ yields the distances of $v$ to every vertex of the graph, *i.e.* a query at a vertex $v$ returns a vector $\vec{v}$, where the $i$th component indicate the distance to $i$th vertex from vertex $v$. It is easy to see that it is a weaker query model as compared to *Layered Graph Query Model*. In the *Distance Query Model*, an edge may be discovered by a combination of queries as illustrated in Figure 2. In the example shown in Fig 2, query at vertex 1 discovers the non-edges $\{(1,4), (1,5), (1,6), (2,6), (3,6)\}$ and edges $\{(1,2), (1,3)\}$. A query at vertex 6 discovers the non-edges
$\{(1,4), (1,5), (1,6), (2,6), (3,6), (4,2), (5,2), (3,2)\}$ and edges $\{(3,1), (1,2), (6,4), (6,5)\}$. Combining these two queries, we discover the edges $(5,3)$ and $(4,3)$. In the off-line version of network discovery problem, the network is initially known to the algorithm. Unlike the online problem, here the goal is to compute a minimum number of queries that suffice to discover the network. Given a network, we can verify whether what we have been given is the correct information. Thus, we refer to the off-line version of network discovery problem as *network verification*.

## 1.1 Prior work in network discovery

Bejerano and Rastogi [5] studied the problem of verifying all edges of a graph with as few queries as possible in a model similar to the *Layered Graph Query Model*. For a graph with $n$ vertices, they give a set-cover based $O(\log n)$-approximation algorithm and show that the problem is NP-hard. In contrast to Bejerano and Rastogi, we are interested in verifying (or discovering) both the edges and the non-edges of a graph. It turns out that the network verification problem was considered as a problem of placing landmarks in graphs [13]. The problem was shown to be NP-complete and an $O(\log n)$-approximation algorithm was presented. Beerliova et al. [3] proved an $\Omega(\log n)$ lower bound on the approximation factor for any polynomial time algorithm for the network verification in the *Layered Graph Query Model* unless $P = NP$.

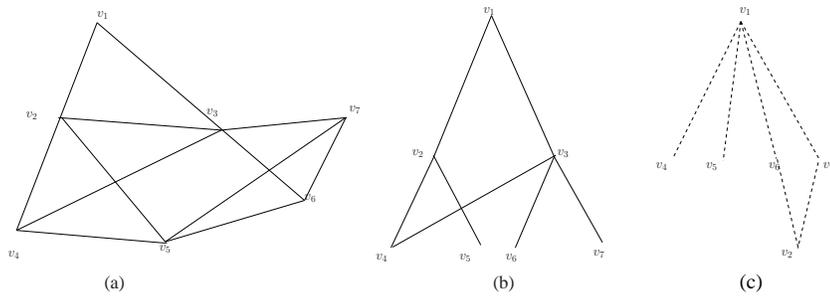

Figure 1: A query at a vertex $v_1$ in the layer graph model (a) yields certificate for the edges in (b) and non-edges in (c)

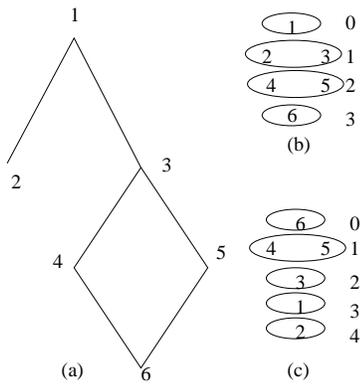

Figure 2: The edges $(5,3)$ and $(4,3)$ of the graph (a) is discovered by the combination of queries at vertex $1$ in (b) and at vertex $6$ in (c) in the distance query model - the distances are depicted via layers of the graph.

In the *online* version of the problem, the network (graph) is unknown to the algorithm. To decide the next query, the algorithm can only use the knowledge about the network it has gained from the answers of previously asked queries. Thus, the difficulty in selecting good queries arises from the fact that we only have the partial information about the network.

For the network discovery problem, Beerliova et al.[4] have shown an $\Omega(\sqrt{n})$ lower bound on the competitive ratio of any deterministic online algorithm and an $\Omega(\log n)$ lower bound for any randomized algorithm for the *Distance Query Model*. The best known algorithm in the *Distance Query Model* is a randomized online algorithm which is $O(\sqrt{n \log n})$-competitive [4]. In contrast, for the *Layered Graph Query Model*, Beerliova et al.[4] have shown that no deterministic online algorithm can be $(3-\varepsilon)$ competitive for any $\varepsilon > 0$. The best known algorithm in this model before this work is an $O(\sqrt{n \log n})$-competitive online randomized algorithm [4] that leaves an exponential gap between the best known lower and upper bounds for the *Layered Graph Query Model*.

In this paper, we present a randomized Monte Carlo online algorithm with a competitive ratio $O(\log^2 n)$ for the *Layered Graph Query Model* thereby nearly closing this exponential gap.

## 1.2 Our results and techniques

The network verification problem can be solved by reducing it to an appropriate instance of the set-cover problem (or hitting set problem). Hence, we obtain an $O(\log n)$ approximation algorithm for the network verification problem which is the best that we can hope to do unless $P = NP$. In the online network discovery problem, we do not know the graph *a priori* and hence the above reduction cannot be used directly. In particular, the sets are not known explicitly, so we first develop an algorithm for solving the covert version of the set-cover problem using queries.

We present an algorithm that computes a set-cover of size at most $O(\log(m' + n') \cdot OPT)$ using at most $O(\log^2(m' + n') \cdot OPT)$ queries with high probability. Using this, we obtain an $O(\log^2 n)$-competitive Monte Carlo randomized algorithm for the network discovery problem in the *Layered Graph Query Model*. This is a significant improvement from the previously best known $O(\sqrt{n \log n})$-competitive algorithm ([3]).

Our algorithm for the set-cover simulates the greedy set-cover algorithm without any information about the contents of any of the sets initially. We use estimation using random sampling to choose the (near) largest cardinality set which is the basis of the greedy algorithm. We have to compensate for the inaccuracies in sampling by using a more careful amortization argument for proving the approximation factor. The greedy algorithm is modified to run in $O(\log(n' + m'))$ rounds instead of the conventional $OPT \cdot \log n'$ stages.

## 2 Preliminaries

Let $G = (V, E)$ be a connected, undirected, unweighted graph representing a network of $n$ vertices. For two distinct nodes $u, v \in V$, we say that $(u, v)$ is an edge if $(u, v) \in E$ and non-edges if $(u, v) \notin E$. The set of non-edges in $G$ is denoted by $\overline{E}$.

We assume that the set $V$ of nodes is known in advance and it is the presence or absence of edges that need to be discovered or verified. A query at node $v$ is denoted by $query(v)$.

We say that a $query(v)$ *certifies* $(u, v)$ if by using the answers to the $query(v)$, one can confirm the presence or absence of the edge $(u, v)$ in the graph, *i.e.* $query(v)$ implicitly or explicitly confirms whether $(u, v) \in E$ or $(u, v) \in \overline{E}$. We associate two sets with each $query(v)$ as follows. For a given vertex $v \in V$, let $Q_v$ denotes the set of all $(u, v) \in V \times V$ such that $query(v)$ *certifies* $(u, v)$. For a given $(u, v) \in V \times V$, let $H_{(u,v)}$ denote the set of all vertices $v$ such that $query(v) certifies$ $(u, v)$. The two definitions can be considered duals of each other.

$$Q_v = \{(u,v) \in V \times V \mid query(v)\ certifies\ (u,v)\}\ \forall v \in V$$

$$H_{(u,v)} = \{v \in V \mid query(v)\ certifies\ (u,v)\} \forall (u,v) \in V \times V.$$

The above formulation of the network discovery problem can be reduced to the *set-cover* problem in which given a collection of sets $Q_v$ of $E \cup \overline{E}$, the goal is to find a (minimum size) subset $V' \subset V$ such that $\cup_{v \in V'} Q_v = E \cup \overline{E}$. Therefore, querying the vertices of the set-cover will certify all the edges and non-edges that can be used to discover the network.

In the related *hitting-set* problem, given a collection of sets $H_{(u,v)}$ of $V$, the goal is to find a (minimum size) subset $V' \subset V$ such that for any given set $H_{(u,v)}$, there exists a vertex $v' \in V'$ such that $v' \in H_{(u,v)}$. It may be noted that the (offline) hitting-set problem is often solved by reducing it to the corresponding set-cover problem.

In the *offline* verification problem, given any query model, one can find the above sets exactly as the graph is known. So the network verification problem can be solved by reducing it to the corresponding set-cover problem (or hitting set problem). Hence, we get an $O(\log n)$ competitive algorithm for the network verification problem. As mentioned earlier this is the best that we can hope to do for this problem unless $P = NP$.

In the online network discovery problem, since we do not know the graph *a priori*, we cannot compute the above sets explicitly without querying all the vertices [2] To circumvent this problem, we develop an algorithm for approximating the set-cover using the related hitting-set queries. It can be easily seen (c.f. Section 6), that that $H_{(u,v)}$ can be obtained from $Q_u$ and $Q_v$ in the context of the network discovery problem.

## 3 Approximating set-cover from $\varepsilon$-net

Clarkson [12] presented an elegant algorithm for set cover for geometric problems with bounded VC dimension (see [10] for a survey of such results) based on *weighted* $\varepsilon$-net. His algorithm is based on random sampling (weighted) and a procedure to verify if a family of subsets is indeed a set cover. Notice that in our context $|C|$ queries suffice to perform this verification where $C$ is the claimed set cover. Intuitively, if an element is covered by $\Omega(\varepsilon)$ fraction of the sets, then it will be covered by the $\varepsilon$-net. For the remaining elements, the algorithms successively boosts the probability of being covered by a clever reweighting technique. The origins of this method goes back to Clarkson [7].

If a set has weight $w$ and the sum of weights of all the sets is $W$, then the set is sampled with probability $\frac{w}{W}$. The algorithm repeatedly picks a random sample where in each new iteration, the weights are modified until we obtain a set cover. The algorithm assumes that the size of the optimal set cover $OPT$ is known and fixes $\varepsilon = \frac{1}{\alpha k}$ where $k = |OPT|$. The algorithm can be summarized as follows

It is known that with high probability, the above algorithm converges in $O(k \log(m/k))$ iterations ([12]). Since $|OPT|$ is not known initially, we can use the doubling technique to guess $|OPT|$ within a factor

---

[2]While this may be necessary for some graphs like the complete graphs, in general this will lead to poor competitive ratio.

**Algorithm 1** Set cover using weighted $\varepsilon$-net
---
Initially assign every set a unit weight. Initialize setcover $C = \phi$.
**while** $C$ is not a cover **do**
1: Pick a weighted $\varepsilon$-net $\mathcal{E}$ of size $O(1/\varepsilon \cdot \log m')$.
2: If $\mathcal{E}$ is a set-cover then report $E$.
3: Else, it misses at least one element, say $x$. Let $S_x$ be the family of all sets that contain the element $x$ and double the weights of all sets in $S_x$.
---

2 by beginning with $k_0 = 1$ and $k_{i+1} = 2k_i$ as the $i$-th guess. Each iteration of the algorithm takes $O(k_i \log m')$ queries, so the total number of queries is $O(k_i^2 \log^2 m')$. Note that it takes $O(k_i)$ queries to verify a set cover. This yields a grand total of $\sum_{i=0}^{\log |OPT|} O(2^{2i} \log^2 m) = O(|OPT|^2 \log^2 m)$ queries. This has competitive ratio roughly $|OPT| \log^2 m$, so that for $|OPT| \leq \tilde{O}(\sqrt{m'})$, the competitive ratio is about $\sqrt{m'}$. For $|OPT| \geq \sqrt{m'}/\log^2 m'$, the competitive ratio is clearly $\tilde{O}(\sqrt{m'})$ as $m'$ queries trivially suffices. Therefore, by using this algorithm for network discovery in the layered graph model, we can match the algorithm of Beerliova et al. [3].

## 4 A near-optimal algorithm

In the conventional greedy set-cover algorithm, we choose a set $\mathbf{s_{max}}$ that covers the maximum number of uncovered elements, say $n_{max}$, and add it to the cover. This leads to a $\log n'$ approximation. Instead, if we choose any set that covers at least half of $n_{max}$ uncovered elements, then it gives a $2 \log n'$ approximation. Recall that $n', m'$ denote the number of elements and the number of sets respectively. More generally, if we choose a set that cover at least $\frac{1}{c} n_{max}$ elements, then we obtain a $c' \log n'$ approximation. We consider a version of this *Relaxed Greedy-Set-Cover* (*RGSC*) where we repeat the following in stages $1, 2, \ldots \log n'$. At any stage we identify all the sets that contain at least $\frac{1}{2} n_{max}$ uncovered elements. We can consider the sets of in an arbitrary, but fixed ordering $O$ and include those sets that contribute at least $\frac{1}{2} n_{max}$ uncovered elements by deleting elements that have been already covered by sets chosen earlier. Note that the sets that will be included will depend on $O$ - however, at the end of this stage, there will not be any set that contains $n_{max}/2$ or more uncovered elements. Since any such ordering $O$ corresponds to a valid run of *RGSC*, this will yield a $2 \log n'$ approximation guarantee - see Appendix for a formal proof.

Our algorithm is based around simulating this approach, where we try to estimate the value of $n_{max}$ indirectly using random sampling. In round $i$, [3] we check for $n_{max} \in [\frac{n'}{2^{i-1}}, \frac{n'}{2^{i-2}}]$ by choosing a random set of uncovered elements of an appropriate size. Using hitting set queries, we find the sets containing these randomly chosen elements. We choose an appropriate number of uncovered elements that will hit the sets having $\frac{n'}{2^{i-2}}$ elements with high probability. We consider the sets in a fixed order and if a set contains more than at least a threshold number of randomly picked elements, then we include the set in the set-cover. Because of the estimation using random sampling, we lose a factor $c' > 2$ in the underlying RGSC as we may choose some sets which contain fewer than $n_{max}/2$ uncovered elements (but at least $\frac{n_{max}}{c'}$).

Algorithm **Pseudo Greedy** described below, selects all sets containing at least $n_{max}/2$ uncovered elements and discards the sets containing less than $\frac{1}{c'} n_{max}$ uncovered elements for $4 < c' < 8$ with high probability.

---
[3]the notation $n_{max}$ will refer to the maximum in the current round $i$.

We assume that the sets are numbered in some canonical order. In the specific application of the network discovery problem, this ordering is implicit ($\{v_1, v_2, \ldots v_n\}$, this induces a canonical ordering on the collection $Q_v$ of sets). In the general setting, we assume that such an ordering exits or it can be easily computed.

In Algorithm 2, $N$ denotes the cardinality of the ground set plus the number sets in the given family ($N = n' + m'$). In the case of Network Discovery problem, $N = O(|V|^2)$. In round $i$, we try to identify the sets containing at least $\frac{n'}{2^{i+1}}$ uncovered elements.

---

**Algorithm 2 Pseudo-Greedy**
---
Initialize set cover $\mathcal{C} = \{\}$.
**for** $i = 0, 1 \ldots$ **do**

1: Let $n_i$ be the number of elements left in this round and $s_i = \min\{\frac{n'}{2^i}, n_i\}$. Choose a random sample $R^i$ of size $(4\alpha n_i/s_i) \log N$.
   **Comment**: $\alpha$ is a constant whose value will be determined in the analysis.

2: If $s_i \leq \alpha \log N$ then solve the hitting set problem directly using at most $n_i$ hitting set queries and run the explicit greedy set-cover algorithm.

3: Else (if $s_i > \alpha \log N$), let $S^i$ be the sets that contain more than $\alpha \log N$ sampled elements. If $S^i$ is empty, increment $i$ and go to step 1.

4: Process $S^i = \{X_1, X_2, \ldots\}$ in some predefined order until all sets are exhausted.

   (i) Let $R_j$ be the union of elements of $R^i$ that are contained in the sets chosen among $X_1, X_2, \ldots X_j$.

   (ii) $\mathcal{C} = \mathcal{C} \cup X_{j+1}$ if
   $$|X_{j+1} \cap (R^i \setminus R_j)| \geq \alpha \log N$$
   (else discard $X_{j+1}$.

   (iii) Update $R_j$ to $R_{j+1}$. using set queries.

5: Update the elements covered by the sets chosen in this round using set queries.

---

## 5 Analysis

We begin with a rough intuition behind the previous algorithm. If the largest set has size $n'/t$ then the minimum number of sets in any set cover is $\Omega(t)$. Therefore we can afford to query a sample of size approximately $O(t \cdot \text{polylog } n')$ elements without blowing up the competitive ratio. In this context note that a uniform random sample of size $O(t \cdot \text{polylog } n')$ will have $\theta(\text{polylog } n')$ elements common with a set of size $n'/t$ with high probability. However, if there are $\Omega(t)$ sets of size $O(n'/t)$, we cannot afford to sample repeatedly for finding these sets. The above observations form the crux of the analysis that are now formalized.

**Lemma 5.1** *In round $i$, in Step 3, the following holds with high probability*
*(i) If a set $T$ contains at least $s_i/2$ elements then with high probability it will have at least $\alpha \log N$ sampled elements.*
*(ii) Any set $T$ chosen in Step 3 will contain at least $\frac{1}{c'}s_i$ elements for $4 < c' < 8$ with high probability.*

*Proof.* Let $T$ be a set where $m \geq |T| \geq m/2$. Suppose we sample every element independently with probability $p$. The expected number of sampled elements $Y$ is such that $mp \geq Y \geq mp/2$. From Chernoff bounds,

$$Pr[(1+\varepsilon)mp \geq Y \geq (1-\varepsilon)mp/2] \geq 1 - 2e^{-mp\varepsilon^2/4}$$

Choosing $\varepsilon = 1/2$, we get

$$Pr[3/2mp \geq Y \geq mp/4] \geq 1 - 2e^{-mp/16}$$

In round $i$, each element is picked independently with probability $(4\alpha/s_i) \log N$, therefore, the expected number of hits in a set of size $m$ is $(4m\alpha/s_i) \log N$. From Chernoff bounds, by substituting $m = s_i$,

$$Pr[6\alpha \log N \geq Y \geq \alpha \log N] \geq 1 - 2e^{-\alpha/4 \log N} = 1 - 2/N^{\alpha/4}$$

Since the number of such $T$ is less than $N$, the algorithm picks all sets containing at least $s_i/2$ uncovered elements with high probability. On the other hand, $T$ be any set chosen in Step 3 of the algorithm. Then, by applying Chernoff bound, we get,

$$Pr[T < s_i/c'] \leq e^{-(c'-2)\alpha/8 \log N} = 1/N^{(c'-2)\alpha/8}$$

for all $4 < c' < 8$. ∎

**Lemma 5.2** *If round $i$ takes $O(\frac{n_i}{s_i} \cdot f(N))$ queries, then the set-cover can be found using $O(n_g \cdot f(N))$ queries where $n_g$ is the size of the set-cover returned by the underlying RGSC Algorithm.*

*Proof.*
In round $i$, we include all those sets in the cover that covers at least $s_i/2$ additional elements. In round $i$, let us distribute the cost uniformly to the remaining elements, i.e., each of the $n_i$ elements is charged $O(f(N)/s_i)$. If an element is covered by a set chosen in round $i$ then it is not charged in the subsequent rounds. So the total cost over all the rounds for element $x$ is $C(x) \leq f(N) \cdot \left(\frac{1}{|s(x)|} + \frac{1}{s_i} + \frac{1}{2s_i} + \frac{1}{4s_i} + \ldots\right) \leq 3c' \frac{f(N)}{|s(x)|}$ where $s(x)$ is the set that *first* covers element $x$ and $s_i/c' \leq |s(x)| \leq s_i$. The constant $c'$ refers to the constant in the previous lemma. Therefore

$$\sum_x C(x) \leq f(N) \sum_x \frac{3c'}{|s(x)|} = 3c'f(N) \cdot \sum_{S \in \mathcal{C}} \sum_{x:s(x)=S} \frac{1}{|s(x)|}$$

The summation represents the cost of the underlying *RGSC* algorithm and therefore, it is bounded by $3c'f(N) \cdot n_g$ (see Lemma 7.1 in the Appendix).
Note that the underlying *RGSC* algorithm is a $c' \log N$ approximation to the set-cover. ∎

**Theorem 5.3** *Algorithm 2 returns a set-cover of size at most $O(\log N \cdot OPT)$ using at most $O(\log^2 N \cdot OPT)$ queries with high probability.*

*Proof.* In our algorithm, $f(N)$ is $O(\log N)$ and the maximum number of iterations is $O(\log N)$. When $s_i < \alpha \log N$, we solve the problem directly using at most $n_i$ hitting set queries, and explicitly run the greedy set-cover. Since the largest set has size $n'/2^i$, the size of the optimal cover is at least $\Omega(n'/\log N)$ and therefore, the number of queries is $O(\log N \cdot OPT)$. In order to prove the theorem, we will show that the bound on $n_g$ in Lemma 5.2 is $O(\log N \cdot OPT)$. So, we must establish that the sets *shortlisted* in Step 3 of the Algorithm and finally included in the cover in Step 4 are only those sets (on the basis of their estimates) that covers at least $s_i/c'$ uncovered elements. In particular, we must guard against oversampling of the uncovered elements of any set *at the time it is considered for inclusion in a given round*. Even though the sets were appropriately sampled in Step 3, at the time of its consideration in Step 4(ii), the sampling of the remaining part must be accurate enough that may necessitate arguing about an exponential number of possibilities depending on the order of its inclusion.

To avoid this, let us assume that we consider the sets of $S^i$ in increasing order of their indices Let $X_1, X_2 \ldots$ be the sets of $S^i$ in this canonical ordering that contain at least $s_i/c'$ elements. Now consider a hypothetical ordering $\mathcal{O}$ of the elements based on this ordering of the sets. Namely, all elements of $X_i$ are numbered smaller than $X_{i+1}$ and the numbering within a set is arbitrary. For example, all the elements in $X_1 \setminus X_2$ are numbered before $X_1 \cap X_2$ and elements of $X_2 \setminus X_1$ come last. Suppose the elements are sampled according to $\mathcal{O}$. We define $X'_i$ as all the uncovered elements in $X_i$ after $X_1, X_2, \ldots X_{i-1}$ have been considered and (hypothetically) sample the elements in $X'_i$ according to $\mathcal{O}$.

We consider $X'_i$ to be *under-sampled* if $|X'_i| \geq s_i$ but the number of sampled elements intersecting $X'_i$ (not including $X_i \setminus X'_i$) is less than $\alpha \log n$. We analogously define *oversampling* for $X'_i$.

We say that a bad event has occurred in round $j$, if any of the sets $X'_i$ is under-sampled or oversampled and let the complement of this event be $Z_i$. From Lemma 5.1, we can bound the probability of under sampling and over sampling such that $\Pr[Z_i] \geq 1 - 3/N^{(\alpha/4)-1}$ (by choosing $c' > 4$). Let $A_i$ be the event that no under-sampling or oversampling occurs for $X'_1, X'_2 \ldots X'_i$. Then,

$$\Pr[A_i] = \Pr[A_{i-1} \bigcap Z_i] = \Pr[Z_i | A_{i-1}] \cdot \Pr[A_{i-1}]$$

Therefore,

$$\Pr[A_i] \geq \Pr[A_{i-1}] \cdot (1 - 3/N^{(\alpha/4)-1}) \geq \left(1 - 2/N^{\alpha/4-1}\right)^i \text{ for } i \leq N$$

By choosing sufficiently large $\alpha$ this is at least $1 - 1/N^2$. Since this holds for all $j \leq O(\log N)$ rounds, this also bounds the failure probability of our algorithm. ∎

**Remarks**: (i) The bounds do not depend on $\mathcal{O}$ and holds for any parallel sampling method.
(ii) We say that the algorithm *fails* if in any of rounds, it does not pick all sets containing at least $s_i/2$ uncovered elements or picks any set containing less than $s_i/c'$ uncovered elements. The sizes of sets that will be chosen will satisfy the the above mentioned bounds with high probability; otherwise, the algorithm will be deemed to have failed. Note that the bound of Lemma 5.2 also holds with the same probability. Since we do not verify these properties, we obtain a Monte Carlo algorithm.
(iii) A deterministic algorithm picks all the sets of size at least $s_i/2$, and while our randomized algorithm chooses all sets of size at least $s_i/2$, it may pick some sets which are little smaller (but greater than $s_i/c'$).

## 6 Network Discovery

The off-line problem of network verification can be reduced to a set-cover problem. In the online version, we do not want to compute the sets explicitly since this will lead to a poor competitive ratio in many situations. So we solve the problem by using hitting-set queries as described in the previous section that gives us an estimate of the set sizes. In our setting, the hitting-set problem is defined on the sets $H_{(u,v)}$ and the set-cover problem on the sets $Q_v$. During any stage, random sampling is done on the set of unresolved edges to obtain estimates of $Q_v$ by querying $Q_{xy}$ where $(x, y)$ is a sampled edge.

Recall that in *Layered Graph Query Model*, a query at a vertex $v$ yields the set of all edges on shortest paths between $v$ and any other vertex. Now, we observe that this query model is equivalent to the model in which a query at vertex $v$ yields all edges and non-edges between vertices of different distances from $v$. Note that an edge connects two vertices of different distance from $v$ if and only if it lies on a shortest path between $v$ and one of these two vertices. The shortest path rooted at $v$ implicitly confirms the absence of all edges between vertices of different distance from $v$. So given an edge or non-edge whose status is not yet resolved, say $(v, u)$, we query both the end points $v$ and $u$ to determine the distances of all nodes to $u$ and $v$. From this we can deduce the set $H_{(u,v)}$ of nodes from which the edge or non-edge between $u$ and $v$ can be discovered: $H_{(u,v)} = \{x \in V | d(u,x) \neq d(v,x) \ d(s,x) = \text{distance from } s \text{ to } x\}$

Algorithm **Pseudo Greedy** described in the previous section above translates to the following in the context of the network discovery problem. Randomly pick a undiscovered edge and query the set $H_{(u,v)}$. Let $n$ be the number of vertices in the graph and let $Q$ denote the query set- this is the (approximately minimal) set of vertices which will be used to discover the network. If $v$ is contained in at least $\alpha \log n$ of the queried sets, include $v$ in the set-cover $Q$. Actually, like the general set-cover problem, it is a two stage process where we first shortlist and then subsequently run through this list in some predefined ordering, say according to the labels of the vertices. As before, we solve the set-cover problem on $Q_v$ using a sequence of $H_{(u,v)}$ hitting set queries. The reader can easily work out the details that we omit to avoid repetition.

In the following algorithm $N = O(n^2)$. The algorithm takes $O(\log n)$ stages and in each stage we make $O(\log n \cdot \text{OPT})$ queries, where OPT is the optimum number of queries required to solve the network verification problem. Since this is also optimum for the online problem, Algorithm **Network Discovery** makes $O(\log^2 n \cdot \text{OPT})$ queries. The algorithm yields a set $O(\log n \cdot \text{OPT})$ $Q_v$ queries that suffices to discover the given network. Therefore the overall number of queries for the online discovery is still $O(\log^2 n \cdot \text{OPT})$.

---

**Algorithm 3** Network Discovery

**for** $i = 0, 1 \ldots$ **do**

1: Let $n_i$ be the number of edges and non-edges which needs to be discovered and $s_i = \min\{\frac{\binom{n}{2}}{2^i}, n_i\}$. Choose a random sample of $R^i$ of size $(4\alpha n_i/s_i) \log N$.
2: If $s_i \leq \alpha \log N$ then find $H_{(u,v)}$ for each of the undiscovered edge/non-edge and solve the network discovery problem by reducing it explicitly to the set-cover problem.
3: If $s_i > \alpha \log N$, for each sampled edge/non-edge $(u, v)$, find the set $H_{(u,v)}$.
4: Consider the vertices $\{v_1, v_2, \ldots\}$ in this order and include $v_j$ in $Q$ ($Q_{v_j}$ is in the set-cover) only if $Q_{v_j}$ contains more than $\alpha \log N$ sampled edge/non-edge. ($v_j \in H_{(u,v)}$ for at least $\alpha \log N$ of the $(u, v) \in R_i$).
   (The actual implementation of this is similar to Steps 3-4 of the Algorithm **Pseudo Greedy**.)

---

From our earlier analysis of the covert set-cover problem it follows that

**Theorem 6.1** *There is a $O(\log^2 n)$-competitive randomized Monte Carlo algorithm for the network discovery problem in the Layered Graph Query Model.*

**Remark** Even if we restrict the query model to return a Layered graph of some bounded depth (that may not correspond to the entire graph), the reduction to covert set cover problem is analogous by modifying the definition of the sets and we obtain the same competitive ratio.

# 7 Conclusion and open problem

The algorithm described in the last section gave a $O(\log^2 n)$ algorithm for the network discovery problem – Can we improve this to $O(\log n)$ ? We can consider a weighted version of the network discovery problem, where each query at a vertex costs say $w_v$, it is not clear whether we can extend our approach to solve the weighted version of the problem.

We note that in the *Distance Query Model*, by querying both $v$ and $u$, we can discover if $u$ or $v$ is a edge or non-edge. If it is a non-edge, then we can find the set $H_{(u,v)}$ – a vertex $w$ is in this set if $d(u,w) - d(v,w) \geq 2$. But if $(u,v)$ is an edge, then we can not find the set $H_{(u,v)}$. It is not clear how to determine the *partial* witnesses, using set-cover queries as before. Therefore, it remains open if we can we improve the known $O(\sqrt{n \log n})$ bound for network discovery problem to $O(poly(\log n))$ approximation randomized algorithm in the *Distance Query Model*?

**Acknowledgement** The first author is thankful to Rajeev Raman and Thomas Erlebach for introducing him to the problem and subsequent technical discussions.

# Appendix A

**Chernoff bounds**

If a random variable $X$ is the sum of $n$ iid Bernoulli trials with a success probability of $p$ in each trial, the following equations give us concentration bounds of deviation of $X$ from the expected value of $np$. These are useful for small deviations from a large expected value.

$$Prob(X \leq (1-\epsilon)pn) \leq exp(-\epsilon^2 np/2) \quad (1)$$

$$Prob(X \geq (1+\epsilon)np) \leq exp(-\epsilon^2 np/4) \quad (2)$$

for all $0 < \epsilon < 1$.

**Greedy set-cover**

For completeness, we also sketch the proof of approximation factor of $RGSC(\theta)$ for $\theta < 1$, such that at any step, the size of the set chosen is at least $\theta \cdot n_{max}$.

Let us number the elements of $S$ in the order they were covered by the greedy algorithm (wlog, we can renumber such that they are $x_1, x_2 \ldots$). We will apportion the cost of covering an element $e \in S$ as $w(e) = \frac{1}{U \setminus V}$ where $e$ is covered for the first time by $U$ and $V$ is set of elements covered till then. This is also called the *cost-effectiveness* of set $U$. The total cost of the cover is

$$\sum_U \sum_{e \in n(U)} \frac{1}{|n(U)|}$$

where $n(U)$ is subset of uncovered elements in $U$ when $U$ was chosen and $e$ is covered for the first time. This can be rewritten as $\sum_i w(x_i)$.

**Lemma 7.1**

$$w(x_i) \leq \frac{C_o/\theta}{n-i+1}$$

*where $C_o$ is the number of sets in the optimum cover.*

In the iteration when $x_i$ is covered for the first time, the number of uncovered elements is $\geq n-i+1$. The pure greedy choice is more cost effective than any left over set of the optimal cover. Suppose $S_{i_1}, S_{i_2} \ldots S_{i_k}$ are the unselected sets of the minimum set-cover. Then, at least one of them has a cost-effectiveness of $\leq \frac{k}{n-i+1} \leq \frac{C_o}{n-i+1}$. It follows that the set chosen by $RGSC(\theta)$ achieves a cost-effectiveness of $\frac{C_o}{(n-i+1)\theta}$. So $w(x_i) \leq \frac{C_o/\theta}{n-i+1}$.

Thus the cost of the greedy cover is $\sum_i \frac{C_o/\theta}{n-i+1}$ which is bounded by $C_o/\theta \cdot H_n$. Here $H_n = \frac{1}{n} + \frac{1}{n-1} + \ldots 1$.